\documentclass[american]{llncs}
\usepackage[T1]{fontenc}
\usepackage[latin9]{inputenc}
\usepackage{babel}
\usepackage{refstyle}
\usepackage{graphicx}
\usepackage[unicode=true]
 {hyperref}

\makeatletter

\AtBeginDocument{\providecommand\figref[1]{\ref{fig:#1}}}
\RS@ifundefined{subsecref}
  {\newref{subsec}{name = \RSsectxt}}
  {}
\RS@ifundefined{thmref}
  {\def\RSthmtxt{theorem~}\newref{thm}{name = \RSthmtxt}}
  {}
\RS@ifundefined{lemref}
  {\def\RSlemtxt{lemma~}\newref{lem}{name = \RSlemtxt}}
  {}

\usepackage{breakcites}

\newref{alg}{refcmd = {Alg.~\ref{#1}}}
\newref{fig}{refcmd = {Fig.~\ref{#1}}}
\newref{tab}{refcmd = {Table~\ref{#1}}}
\newref{sec}{refcmd = {\bf "\ref{#1}"}}
\newref{sub}{refcmd = {\bf"\ref{#1}"}}
\newref{apx}{refcmd = {Appendix~\ref{#1}}}
\newref{eq}{refcmd = {Eq.~(\ref{#1})}}
\newref{eqs}{refcmd = {Eqs.~(\ref{#1})}}

\makeatother

\usepackage{listings}

\begin{document}
\title{Usage and Scaling of an Open-Source Spiking Multi-Area Model of Monkey
Cortex}
\author{S.J. van Albada$^{1,2}$, J. Pronold$^{1,3}$, A. van Meegen$^{1,2}$,
and M. Diesmann$^{1,4,5}$}
\institute{$^{1}$Institute of Neuroscience and Medicine (INM-6), Institute for
Advanced Simulation (IAS-6), and JARA Institute Brain Structure-Function
Relationships (INM-10), Jülich Research Centre, Jülich, Germany\\
$^{2}$Institute of Zoology, Faculty of Mathematics and Natural Sciences,
University of Cologne, Germany\\
$^{3}$RWTH Aachen University, Aachen, Germany \\
$^{4}$Department of Physics, Faculty 1, RWTH Aachen University, Aachen,
Germany \\
$^{5}$Department of Psychiatry, Psychotherapy and Psychosomatics,
School of Medicine, RWTH Aachen University, Aachen, Germany}
\maketitle
\begin{abstract}
We are entering an age of `big' computational neuroscience, in which
neural network models are increasing in size and in numbers of underlying
data sets. Consolidating the zoo of models into large-scale models
simultaneously consistent with a wide range of data is only possible
through the effort of large teams, which can be spread across multiple
research institutions. To ensure that computational neuroscientists
can build on each other's work, it is important to make models publicly
available as well-documented code. This chapter describes such an
open-source model, which relates the connectivity structure of all
vision-related cortical areas of the macaque monkey with their resting-state
dynamics. We give a brief overview of how to use the executable model
specification, which employs NEST as simulation engine, and show its
runtime scaling. The solutions found serve as an example for organizing
the workflow of future models from the raw experimental data to the
visualization of the results, expose the challenges, and give guidance
for the construction of ICT infrastructure for neuroscience.
\end{abstract}


\section{Introduction}

With the availability of ever more powerful supercomputers, simulation
code that can efficiently make use of these resources \cite{Jordan18_2},
and large, systematized data sets on brain architecture, connectivity,
neuron properties, genetics, transcriptomics, and receptor densities
\cite{vanAlbada2020_arXiv,Zilles02_587,Bakker12_30,Reimann15_1,Ero18_84,Tasic18_72,Gouwens19_1182,Sugino19_e38619,Winnubst19_268},
the time is ripe for creating large-scale models of brain circuitry
and dynamics.

We recently published a spiking model of all vision-related areas
of macaque cortex, relating the network structure to the multi-scale
resting-state activity \cite{Schmidt18_1409,Schmidt18_e1006359}.
The model simultaneously accounts for the parallel spiking activity
of populations of neurons and for functional connectivity as measured
with resting-state functional magnetic resonance imaging (fMRI). As
a spiking network model with the full density of neurons and synapses
in each local microcircuit, yet covering a large part of the cerebral
cortex, it is unique in bringing together realistic microscopic and
macroscopic activity.

Rather than as a finished product, the model is intended as a platform
for further investigations and developments, for instance to study
the origin of oscillations \cite{Shimoura19_CNS}, to add function
\cite{Korcsak-Gorzo19_Bernstein}, or to go toward models of human
cortex \cite{Pronold19_NESTconf}.

To support reuse and further developments by others we have made the
entire executable workflow available, from anatomical data to analysis
and visualization. Here we provide a brief summary of the model, followed
by an overview over the workflow components, along with a few typical
usage examples.

The model is implemented in NEST \cite{Gewaltig_07_11204} and can
be executed using a high-performance compute (HPC) cluster or supercomputer.
We provide corresponding strong scaling results to give an indication
of the necessary resources and optimal parallelization.

\section{Overview of the multi-area model}

The multi-area model describes all 32 vision-related areas in one
hemisphere of macaque cortex in the parcellation of Felleman and Van
Essen \cite{Felleman91_1}. Each area is represented by a layered
spiking network model of a $1\:\mathrm{mm^{2}}$ microcircuit \cite{Potjans14_785},
adjusted to the area- and layer-specific numbers of neurons and laminar
thicknesses. Layers 2/3, 4, 5, and 6 each have an excitatory and an
inhibitory population of integrate-and-fire neurons. To minimize downscaling
distortions \cite{Albada15}, the local circuits contain the full
density of neurons and synapses. This brings the total size of the
network to \ensuremath{\sim}4 million neurons and \ensuremath{\sim}24
billion synapses. All neurons receive an independent Poisson drive
to represent the non-modeled parts of the brain.

\begin{figure}
\includegraphics[width=1\textwidth]{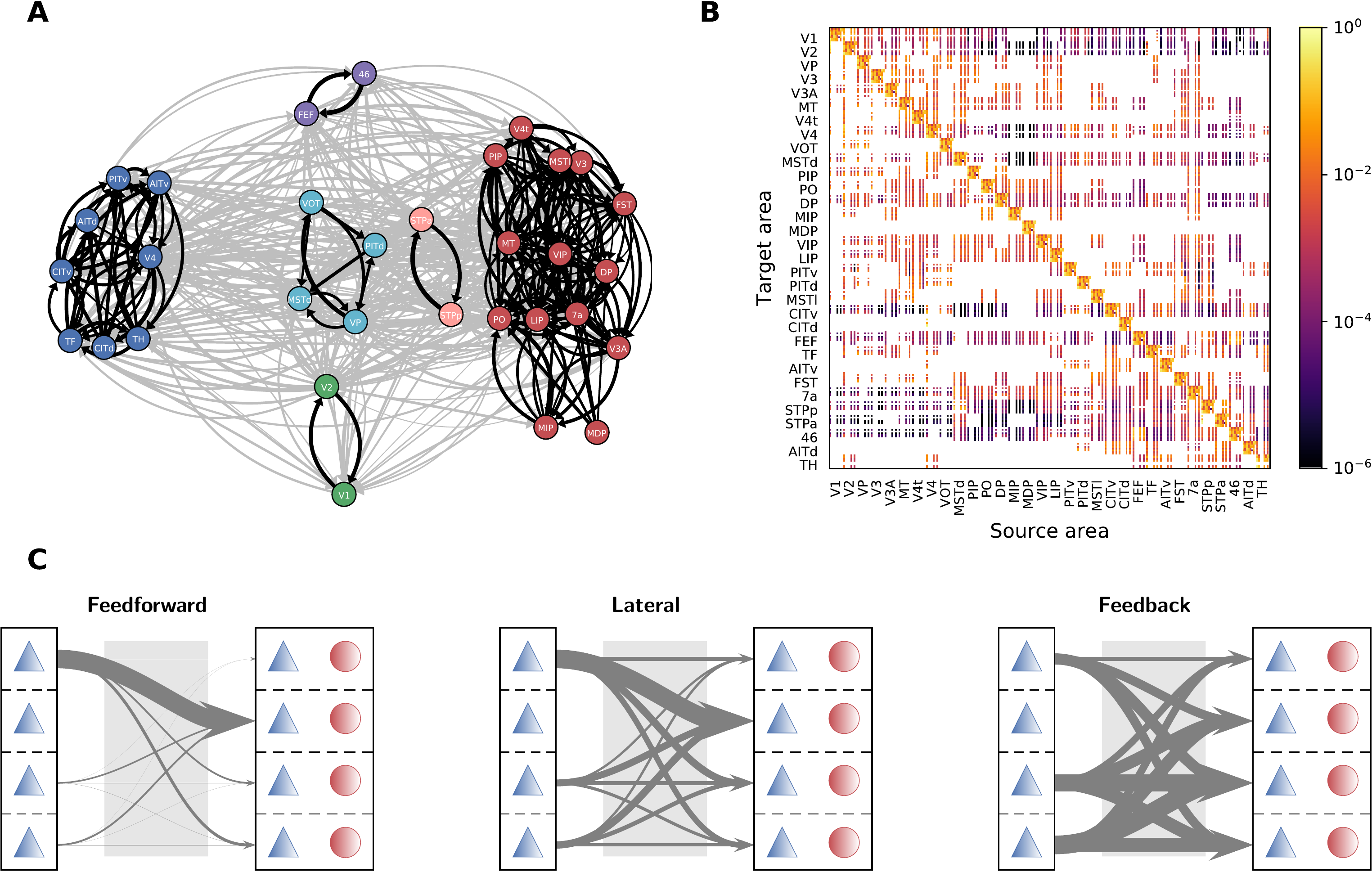}\caption{Overview of the connectivity of the multi-area model as determined
from anatomical data and predictive connectomics. \textbf{A} Area-level
connectivity. The weighted and directed graph of the number of outgoing
synapses per neuron (out-degrees) between each pair of areas is clustered
using the map equation method \cite{Rosvall09}. Green, early visual
areas; dark blue, ventral stream areas; purple, frontal areas; red,
dorsal stream areas; light red, superior temporal polysensory areas;
light blue, mixed cluster. Black arrows show within-cluster connections,
gray arrows between-cluster connections. \textbf{B} Population-level
connection probabilities (the probability of at least one synapse
between a pair of neurons from the given populations). \textbf{C}
Hierarchically resolved average laminar patterns of the numbers of
incoming synapses per neuron (in-degrees). Hierarchical relationships
are defined based on fractions of supragranular labeled neurons ($SLN$)
from retrograde tracing experiments \cite{Markov14}: feedforward,
$SLN>0.65$; lateral, $0.35\protect\leq SLN\protect\leq0.65$; feedback,
$SLN<0.35$. The connections emanate from excitatory neurons and are
sent to both excitatory and inhibitory neurons. For further details
see \cite{Schmidt18_1409}.}
\label{fig:mam_overview}
\end{figure}

The inter-area connectivity is based on axonal tracing data from the
CoCoMac database on the existence and laminar patterns of connections
\cite{Bakker12_30}, along with quantitative tracing data also indicating
the numbers of source neurons in each area and their supragranular
or infragranular location \cite{Markov11_1254,Markov2014_17}. These
data are complemented with statistical predictions (`predictive connectomics')
to fill in the missing values, based on cortical architecture (neuron
densities, laminar thicknesses) and inter-area distances \cite{Hilgetag19_905}.
\figref{mam_overview} shows the resulting connectivity at the level
of areas, layers, and populations. A semi-analytical mean-field method
adjusts the data-based connectivity slightly in order to bring the
firing rates into biologically plausible ranges \cite{Schuecker17}.

By increasing the synaptic strengths of the inter-area connections,
slow activity fluctuations, present in experimental recordings but
not in the isolated microcircuit, are reproduced. In particular, the
system needs to be poised right below an instability between a low-activity
and a high-activity state in order to capture the experimental observations.
The spectrum of the fluctuations and the distribution of single-neuron
spike rates in primary visual cortex (V1) are close to those in lightly
anesthetized macaque monkeys. At the same synaptic strengths where
the parallel spiking activity of V1 neurons is most realistic, also
the inter-area functional connectivity is most similar to macaque
fMRI resting-state functional connectivity.

\section{The multi-area model workflow}

\begin{figure}
\includegraphics[width=1\textwidth]{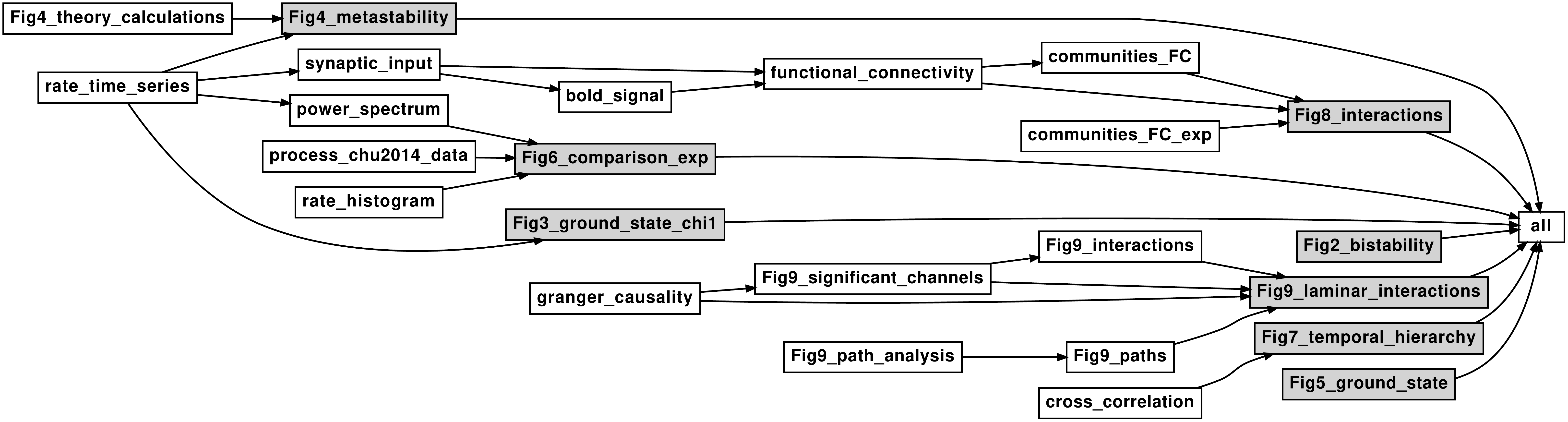}

\caption{\textbf{Visualization of a Snakemake workflow}. The interdependencies
between the scripts reproducing the figures in \cite{Schmidt18_e1006359},
visualized as a directed acyclic graph. The label of a node corresponds
to the name of the script. \label{fig:snakemake}}
\end{figure}

The multi-area model code is available via \href{https://inm-6.github.io/multi-area-model/}{https://inm-6.github.io/multi-area-model/}
and covers the full digitized workflow from the raw experimental data
to simulation, analysis, and visualization. The model can thus be
cloned to obtain a local version, or forked to build on top of it.
The implementation language is Python, the open-source scripting language
the field of computational neuroscience has agreed on \cite{Muller15_11}.
The online documentation provides all information needed to instantiate
and run the model. The tool Snakemake \cite{Koester12_2520} is used
to specify the interdependencies between all the scripts and execute
them in the right order to reproduce the figures of the papers on
the model's anatomy \cite{Schmidt18_1409}, dynamics \cite{Schmidt18_e1006359},
and stabilization based on mean-field theory \cite{Schuecker17} (see
\figref{snakemake}).

Furthermore, if one of the files in the workflow is adjusted, Snakemake
enables executing only that file and the ones that depend on it anew.
A tutorial video (\href{https://www.youtube.com/watch?v=YsH3BcyZBcU}{https://www.youtube.com/watch?v=YsH3BcyZBcU})
gives a brief overview of the model, explains the structure of the
code, and shows how to run a basic simulation.

\section{Example usage}

One main property delivered by the multi-area model is the population-,
\mbox{layer-,} and area-specific connectivity for all vision-related
areas in one hemisphere of macaque cortex. We here describe how to
obtain the two available versions of this connectivity: 1) based directly
on the anatomical data plus predictive connectomics that fills in
the missing values; 2) after slight adjustments in the connectivity
in order to arrive at plausible firing rates in the model. We also
refer to this procedure as `stabilization' because obtaining plausible
activity entails enhancing the size of the basin of attraction of
the low-activity state, i.e., increasing its global stability. An
example how to use the mean-field method developed in \cite{Schuecker17}
for this purpose is provided in the model repository: \href{https://github.com/INM-6/multi-area-model/blob/master/figures/SchueckerSchmidt2017/stabilization.py}{figures/SchueckerSchmidt2017/stabilization.py}.
The method adjusts the number of incoming connections per neuron (in-degree).
The script exports the adjusted matrix of all in-degrees as a NumPy
\cite{Oliphant07} file; to have the matrix properly annotated one
can instantiate the \lstinline[language=Python,basicstyle={\ttfamily},tabsize=4]!MultiAreaModel!
class with the exported matrix specified in the connection parameters
as \lstinline[language=Python,basicstyle={\ttfamily},tabsize=4]!K_stable!.
Afterwards, one can access the connectivity using the instantiation
\lstinline[language=Python,basicstyle={\ttfamily},tabsize=4]!M! of
the \lstinline[language=Python,basicstyle={\ttfamily},tabsize=4]!MultiAreaModel!
class: \lstinline[language=Python,basicstyle={\ttfamily},tabsize=4]!M.K!
for the in-degrees or \lstinline[language=Python,basicstyle={\ttfamily},tabsize=4]!M.synapses!
for the total number of synapses between each pair of populations.
To obtain the connectivity without stabilization, it is sufficient
to instantiate the \lstinline[language=Python,basicstyle={\ttfamily},tabsize=4]!MultiAreaModel!
class without specifying \lstinline[language=Python,basicstyle={\ttfamily},tabsize=4]!K_stable!.

\begin{figure}
\includegraphics[width=1\textwidth]{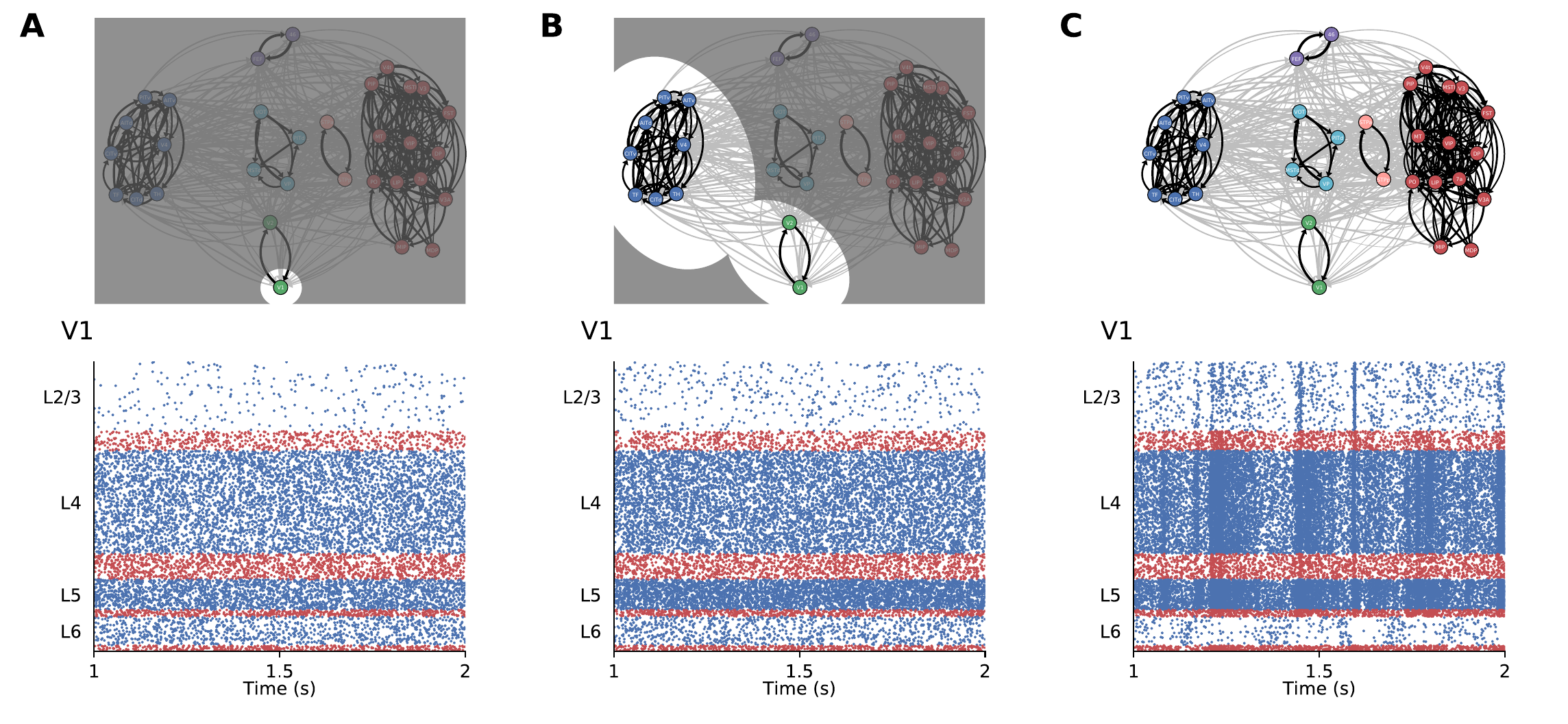}

\caption{\textbf{Simulating subsets of the multi-area model.} Bottom panels,
spiking activity of excitatory (blue) and inhibitory (red) neurons
in all layers of area V1 where in \textbf{A} and \textbf{B} a subset
of areas is replaced by Poisson processes. Top panels, sketches visualizing
which areas were simulated (white spotlights); the colors therein
correspond to different clusters: lower visual (green), ventral stream
(dark blue), dorsal stream (red), superior temporal polysensory (light
red), mixed cluster (light blue), and frontal (purple). If all areas
besides V1 are replaced by Poisson processes (\textbf{A}), the activity
displays no temporal structure. Simulating a subset of ten areas (\textbf{B})
slightly increases the activity but does not give rise to a temporal
structure, either. Only the simulation of the full model (\textbf{C})
produces a clear temporal structure in the activity. Parameters identical
to \cite[Fig. 5]{Schmidt18_e1006359}. \label{fig:partial_simulation}}
\end{figure}

Performing a simulation of the full multi-area model requires a significant
amount of compute resources. To allow for smaller simulations, it
is possible to simulate only a subset of the areas. In this case,
the non-simulated areas can be replaced by Poisson processes with
a specified rate. To this end, the options \lstinline[language=Python,basicstyle={\ttfamily},tabsize=4]!replace_non_simulated_areas!
and \lstinline[language=Python,basicstyle={\ttfamily},tabsize=4]!replace_cc_input_source!
in \lstinline[language=Python,basicstyle={\ttfamily},tabsize=4]!connection_params!
have to be set to \lstinline[language=Python,basicstyle={\ttfamily},tabsize=4]!`het_poisson_stat'!
and the path to a JSON file containing the rates of the non-simulated
areas. Lastly, the simulated areas have to be specified as a list,
for instance
\begin{lstlisting}[language=Python,basicstyle={\ttfamily},showstringspaces=false,tabsize=4,escapeinside={@*}{*@}]
sim_params[`areas_simulated'] = [`V1', `V2']@*\textnormal{,}*@
\end{lstlisting}
 before the \lstinline[language=Python,basicstyle={\ttfamily},tabsize=4]!MultiAreaModel!
class is instantiated. A simple example how to deploy a simulation
is given in \href{https://github.com/INM-6/multi-area-model/blob/master/run_example_fullscale.py}{run\_example\_fullscale.py};
the effect of replacing areas by Poisson processes is shown in \figref{partial_simulation}
$\!$.

\section{Strong scaling}

\begin{figure}
\includegraphics[width=1\textwidth]{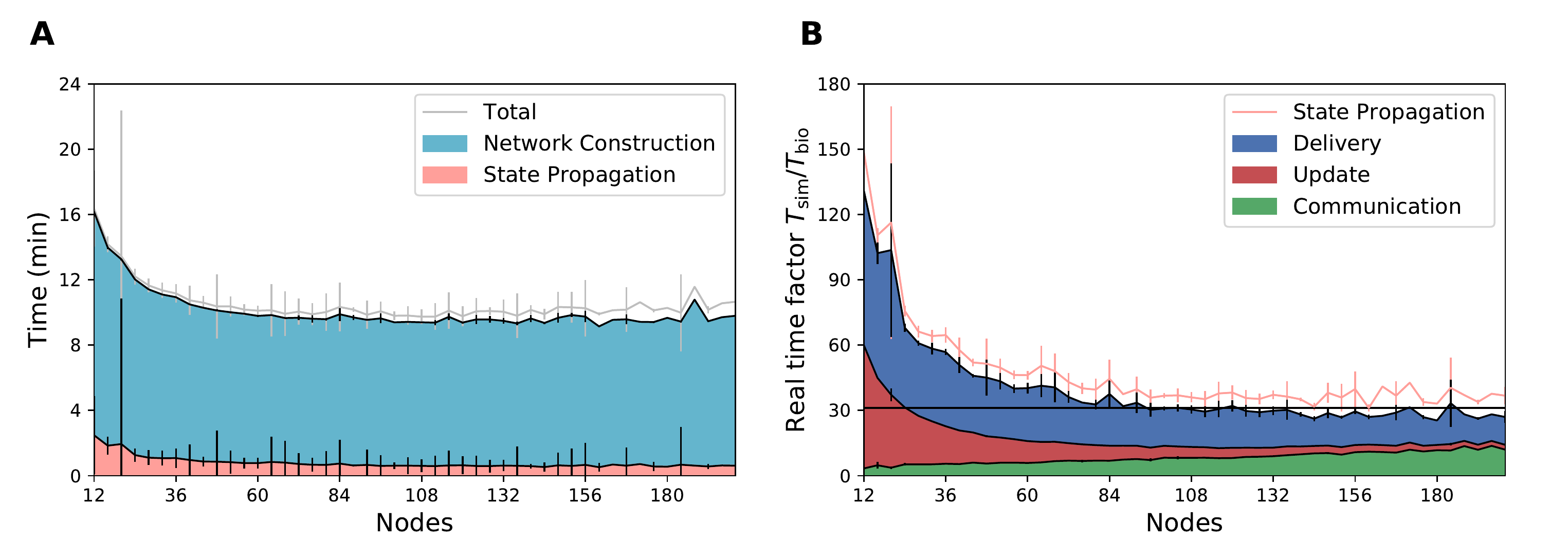}

\caption{\textbf{Strong scaling of the multi-area model}. The contributions
of different phases to the total simulation time and state propagation
for 12 to 200 compute nodes for 1$\:$s of biological time. Each data
point consists of three random network realizations, with error bar
showing the standard deviation of measurements. (\textbf{A}) The main
contributions to the total time are network construction and state
propagation. (\textbf{B}) The state propagation is dominated by three
phases: the communication, update, and spike delivery phase. Adding
more compute resources while keeping network size the same (strong
scaling) decreases the latter two but increases the absolute and
relative contribution of the communication phase. The combined contributions
are minimized at a real-time factor of 31 (black horizontal line).
Parameters identical to \cite[Fig. 5]{Schmidt18_e1006359}. \label{fig:strong_scaling}}
\end{figure}

The limiting factor dictating the necessary compute resources for
simulating the multi-area model is the available memory. Approximately
1$\:$TB is needed for instantiating the network alone. To ensure
sufficient memory for the model, the simulation has to be distributed
across multiple nodes.

We simulated the model using NEST 2.14 \cite{Nest2140} on the JURECA
supercomputer in Jülich, which provides 1872 compute nodes equipped
with two Intel Xeon E5-2680 v3 Haswell CPUs per node. Each CPU has
12 cores running at 2.5$\:$GHz. The standard compute node provides
128$\:$GB of memory of which 96$\:$GB are guaranteed to be available
for the running application. Thus, on this machine, the multi-area
model needs at least 11 nodes.

Having established the minimal hardware requirements, we are interested
in the runtime of the simulation depending on the compute resources.
We quantify the proximity of the time required for state propagation
to biological real time by the real-time factor $T_{\mathrm{sim}}/T_{\mathrm{bio}}$.
Carrying out a strong scaling experiment, the problem size stays fixed
and we increase the number of compute nodes from 12 to 200, thus reducing
the load per node. In our simulations we use 6 MPI processes per node
and 4 threads per MPI process, as we found this hybrid parallelization
to perform better than other combinations of threading and MPI parallelism
(data not shown). In particular during the construction phase, hybrid
parallelization outperformed pure threading by a large margin. The
threads are pinned to the cores and jemalloc is used as a memory allocator
(\cite{Evans2011_fb}, see \cite{Ippen2017_30} for the relevance
of the allocator for NEST). In each run, we simulate 10$\:$s of biological
time.

In \figref{strong_scaling}\textbf{A} the total runtime and its main
contributions, network construction and state propagation, are shown.
The contribution of state propagation is averaged to 1$\:$s of biological
time. The main share of the time is taken up by network construction.
During this phase the neurons and synapses are created and connected,
while during state propagation the dynamics of the model is simulated.
The time spent in the former phase is fixed, as it is independent
of the specified biological time, whereas the time spent propagating
the state depends on the specified biological time and the state of
the network. Depending on the initial conditions and the random Poisson
input, the network exhibits higher or lower activity, affecting the
time spent propagating the state. Hence, the ratio of both phases
should not be taken at face value. In some cases, longer simulations
are of interest, increasing the relevance of the time spent propagating
the state. Thus, it is interesting to know how different components
of the state propagation algorithm contribute to this phase.

The three main phases during state propagation are: update of neuronal
states, communication between MPI processes, and delivery of spikes
to the target neurons within each target MPI process. \figref{strong_scaling}\textbf{B}
shows the contributions of these phases to the real-time factor. Adding
more compute resources brings down the contributions of the update
and delivery phases and increases the time consumption of communication.
Especially the delivery of spikes is heavily dependent on the network
activity. At 160 nodes, a real-time factor of 31 is achieved (mean
spike rate $14.1\,$spikes/s). This slowdown compared to real time
enables researchers to study the dynamics of the multi-area model
over sufficiently long periods, for example in detailed correlation
analyses, but systematic investigations of plasticity and learning
would still profit from further progress.

In order to test the influence of the communication rate on the time
required for state propagation, we carried out a simulation of a two-population
balanced random network \cite{Brunel00} which has been used in previous
publications on neuronal network simulation technology \cite{Helias12_26,Kunkel14_78,Kunkel2017_11,Kunkel2012_5_35,Ippen2017_30,Jordan18_2}.
We use the same parameters as in \cite{Jordan18_2}, but replace connections
governed by spike-timing dependent plasticity by static connections.
In addition we set the numbers of neurons and synapses to match those
in the multi-area model (resulting mean spike rate $12.9\,$spikes/s).
The communication interval is determined by the minimum delay, as
the spikes can be buffered over the duration of this delay while maintaining
causality \cite{Morrison05a}. The multi-area model has a minimum
delay of 0.1$\:$ms, whereas the balanced random network has a uniform
delay of 1.5$\:$ms, so that communication occurs 15 times less often.
Using 160 nodes and the same configurations as before, we find a real-time
factor of 17. Here, 80\% of the time is spent delivering the spikes
to the target neurons locally on each process, whereas only 1\% of
the time is spent on MPI communication. Forcing the two-population
model to communicate in 0.1$\:$ms intervals by adding a synapse of
corresponding delay and zero weight indeed requires the same absolute
time for MPI communication as in the multi-area model. The real-time
factor increases to 34, significantly larger than for the multi-area
model. The increase is entirely due to a longer spike delivery phase.
How the efficiency of spike delivery is determined by the network
activity remains to be answered by future investigations. Possibly
relevant factors are the wide distribution of spike rates in the multi-area
model compared to the narrow one in the two-population model, and
the different synchronization patterns of neuronal activity in the
two models. In summary, less frequent MPI communication shifts the
bottleneck to another software component while almost halving the
total runtime. This opens up the possibility of speeding up the simulation
even more through optimized algorithms for spike delivery on each
target process.

\section{Conclusions}

The usefulness of large-scale data-driven brain models is often questioned
\cite{Eliasmith14_1,Fregnac17_6362,Bassett18_566}, as their high
complexity limits ready insights into mechanisms underlying their
dynamics, large numbers of parameters and a lack of testing of models
with new data may lead to overfitting and poor generalization, and
function does not emerge magically by putting the microscopic building
blocks together. However, this argument can also be turned around.
It seems that in recent years the complexity of the majority of models
and thereby their scope of explained brain functions is not increasing
anymore. One reason is that elegant publications on minimal models
explaining a single phenomenon are often also end points in that they
have no explanatory power beyond their immediate scope. It remains
unclear how the proposed mechanisms interact with other mechanisms
realized in the same brain structure, and how such models can be used
as building blocks for larger models giving a more complete picture
of brain function. The powerful approach of minimal models from physics
needs to be integrated with the systems perspective of biology. To
achieve models able to make accurate predictions for a broad range
of questions, the zoo of available models of individual brain regions
and hypothesized mechanisms needs to be consolidated into large-scale
models tested on numerous benchmarks on multiple scales \cite{Einevoll19_735}.
Having an accurate, if complex, model of the brain that generates
reliable predictions enables\textit{ in silico} experiments, for instance
to predict treatment outcomes for neurological conditions, potentially
even for individual subjects \cite{Proix17_641}. Furthermore, combining
the bottom-up, data-driven approach with a top-down, functional approach
allows models to be equipped with information processing capabilities.
Creating such accurate, integrative models will require overcoming
the complexity barrier computational neuroscience is facing. Without
progress in the software tools supporting collaborative model development
and the expressive digital representation of models and the required
workflows, reproducibility and reusability cannot be maintained for
more complex models.

On the technical side, simulation codes like NEST have matured to
generic simulation engines for a wide range of models. Recent developments
in the simulation technology of NEST have considerably sped up the
state propagation and reduced the memory footprint \cite{Jordan18_2}
of large-scale network models. The rapid state propagation causes
the network construction phase to take up a large fraction of the
simulation time for simulations of short to medium duration. Furthermore,
the fact that hybrid parallelization currently performs better than
pure threading during the construction phase indicates that the code
still spends time on the Python interpreter level and does not yet
optimally make use of memory locality. For these reasons, speeding
up network construction should be a focus of future work.

Our strong scaling results show that communication starts to dominate
at an intermediate number of nodes, so that the further speed-up in
the solution of the neuron equations cannot be fully exploited. Therefore,
it would be desirable to develop methods for further limiting the
time required for communication, for instance by distributing the
neurons across the processes according to the modular structure of
the neuronal network, as opposed to the current round-robin distribution.
The longer delays between areas compared to within areas would then
allow less frequent MPI communication, by buffering the spikes for
the duration of the delay \cite{Morrison05a}. A major fraction of
time is then spent in the spike delivery phase. Here an algorithm
needs to transfer the spikes arriving at the compute node to their
target neurons. It is our hope that in future a better understanding
of the interplay between the intrinsically random access pattern and
memory architecture will lead to more effective algorithms.

While the publication of the model code in a public repository enables
downloading and executing the code, this requires setting up the simulation
on the chosen HPC system, which may be nontrivial, and the HPC resources
have to be available to the research group in the first place. Therefore,
it would be desirable to link computing resources to the repository,
enabling the code to be executed directly from it. The ICT infrastructure
for neuroscience (EBRAINS) being created by the European Human Brain
Project (HBP) has made first steps in this direction. A preliminary
version of a digital workflow for the collaborative development of
reproducible and reusable models was evaluated in \cite{Senk17_243}.
Next to finding a concrete solution for the multi-area model at hand,
the purpose of the present study was to extend the previous work and
obtain a clearer picture of the requirements on collaborative model
development and the digital representation of workflows. From the
present perspective it seems effective not to reimplement the functionality
of advanced code development platforms like GitHub in the HBP infrastructure
but to build a bridge enabling execution of the models and storage
of the results. An essential feature will be that the model repository
remains portable by abstractions from any machine specific instructions
and authorization information.

The microcircuit building block for this model \cite{Potjans14_785}
has found strong resonance in the computational neuroscience community,
having already inspired multiple follow-up studies \cite{Linden11_859,Cain16_e1005045,Hagen16,Schwalger17_e1005507,Shimoura18_ReScience,VanAlbada18_291,Knight18_941,Rhodes19_20190160}.
The multi-area model of monkey cortex developed by Schmidt et al.~\cite{Schmidt18_1409,Schmidt18_e1006359}
and described here has a somewhat higher threshold for reuse, due
to its greater complexity and specificity. Nevertheless, it has already
been ported to a single GPU using connectivity generated on the fly
each time a spike is triggered, thereby trading memory storage and
retrieval for computation, which is possible in this case because
the synapses are static \cite{Knight2020_bioRxiv}. We hope that the
technologies presented here push the complexity barrier of neuroscience
modeling a bit further out, such that the model will find a wide uptake
and serve as a scaffold for generating an ever more complete and realistic
picture of cortical structure, dynamics, and function.

\section*{Acknowledgments}

Supported by the European Union's Horizon 2020 research and innovation
program under HBP SGA1 (Grant Agreement No. 720270), the European
Union\textquoteright s Horizon 2020 Framework Programme for Research
and Innovation under Specific Grant Agreements No. 785907 and 945539
(Human Brain Project SGA2, SGA3), Priority Program 2041 (SPP 2041)
\textquotedbl Computational Connectomics\textquotedbl{} of the German
Research Foundation (DFG), and the Helmholtz Association Initiative
and Networking Fund under project number SO-092 (Advanced Computing
Architectures, ACA). Simulations on the JURECA supercomputer at the
Jülich Supercomputing Centre were enabled by computation time grant
JINB33.


\begin{thebibliography}{10}
\bibitem{Jordan18_2} Jordan, J., Ippen, T., Helias, M., Kitayama, I., Sato, M., Igarashi, J., Diesmann, M., Kunkel, S.: \newblock Extremely scalable spiking neuronal network simulation code: From   laptops to exascale computers. \newblock Front. Neuroinform. \textbf{12} (2018) ~2
\bibitem{vanAlbada2020_arXiv} van Albada, S.J., Morales-Gregorio, A., Dickscheid, T., Goulas, A., Bakker, R., Bludau, S., Palm, G., Hilgetag, C.-C., Diesmann, M.: \newblock Bringing anatomical information into neuronal network models. \newblock arXiv preprint arXiv:2007.00031 (2020)
\bibitem{Zilles02_587} Zilles, K., Palomero-Gallagher, N., Grefkes, C., Scheperjans, F., Boy, C., Amunts, K., Schleicher, A.: \newblock Architectonics of the human cerebral cortex and transmitter receptor   fingerprints: reconciling functional neuroanatomy and neurochemistry. \newblock Eur. Neuropsychopharmacol. \textbf{12}(6) (2002)  587--599
\bibitem{Bakker12_30} Bakker, R., Thomas, W., Diesmann, M.: \newblock {CoCoMac} 2.0 and the future of tract-tracing databases. \newblock Front. Neuroinform. \textbf{6} (2012) ~30
\bibitem{Reimann15_1} Reimann, M.W., King, J.G., Muller, E.B., Ramaswamy, S., Markram, H.: \newblock An algorithm to predict the connectome of neural microcircuits. \newblock Front. Comput. Neurosci. \textbf{9} (2015)  120
\bibitem{Ero18_84} Er{\"o}, C., Gewaltig, M.O., Keller, D., Markram, H.: \newblock A cell atlas for the mouse brain. \newblock Front. Neuroinform. \textbf{12} (2018) ~84
\bibitem{Tasic18_72} Tasic, B., Yao, Z., Graybuck, L.T., Smith, K.A., Nguyen, T.N., Bertagnolli, D.,   Goldy, J., Garren, E., Economo, M.N., Viswanathan, S.,  et~al.: \newblock Shared and distinct transcriptomic cell types across neocortical   areas. \newblock Nature \textbf{563} (2018)  72--78
\bibitem{Gouwens19_1182} Gouwens, N.W., Sorensen, S.A., Berg, J., Lee, C., Jarsky, T., Ting, J., Sunkin, S.M., Feng, D., Anastassiou, C.A., Barkan, E.,  et~al.: \newblock Classification of electrophysiological and morphological neuron types   in the mouse visual cortex. \newblock Nat. Neurosci. \textbf{22} (2019)  1182--1195
\bibitem{Sugino19_e38619} Sugino, K., Clark, E., Schulmann, A., Shima, Y., Wang, L., Hunt, D.L., Hooks, B.M., Tr{\"a}nkner, D., Chandrashekar, J., Picard, S.,  et~al.: \newblock Mapping the transcriptional diversity of genetically and anatomically defined cell populations in the mouse brain. \newblock eLife \textbf{8} (2019)  e38619
\bibitem{Winnubst19_268} Winnubst, J., Bas, E., Ferreira, T.A., Wu, Z., Economo, M.N., Edson, P.,   Arthur, B.J., Bruns, C., Rokicki, K., Schauder, D.,  et~al.: \newblock Reconstruction of 1,000 projection neurons reveals new cell types and   organization of long-range connectivity in the mouse brain. \newblock Cell \textbf{179}(1) (2019)  268--281
\bibitem{Schmidt18_1409} Schmidt, M., Bakker, R., Hilgetag, C.C., Diesmann, M., van Albada, S.J.: \newblock Multi-scale account of the network structure of macaque visual   cortex. \newblock Brain Struct. Func. \textbf{223}(3) (2018)  1409--1435
\bibitem{Schmidt18_e1006359} Schmidt, M., Bakker, R., Shen, K., Bezgin, G., Diesmann, M., van Albada, S.J.: \newblock A multi-scale layer-resolved spiking network model of resting-state   dynamics in macaque visual cortical areas. \newblock PLOS Comput. Biol. \textbf{14}(10) (2018)  e1006359
\bibitem{Shimoura19_CNS} Shimoura, R.O., Roque, A.C., Diesmann, M., van Albada, S.J.: \newblock Visual alpha generators in a spiking thalamocortical microcircuit   model. \newblock In: {28th Annual Computational Neuroscience Meeting}. (2019)  P204
\bibitem{Korcsak-Gorzo19_Bernstein} Korcsak-Gorzo, A., van Meegen, A., Scherr, F., Subramoney, A., Maass, W., van Albada, S.J.: \newblock Learning-to-learn in data-based columnar models of visual cortex. \newblock In: {Bernstein Conference 2019}. (2019) ~W9
\bibitem{Pronold19_NESTconf} Pronold, J., van Meegen, A., Bakker, R., Morales-Gregorio, A., van Albada,   S.J.: \newblock Multi-area spiking network models of macaque and human cortices. \newblock In: {NEST Conference 2019}. (2019) ~30
\bibitem{Gewaltig_07_11204} Gewaltig, M.O., Diesmann, M.: \newblock {NEST} ({NE}ural {S}imulation {T}ool). \newblock Scholarpedia \textbf{2}(4) (2007)  1430
\bibitem{Felleman91_1} Felleman, D.J., {Van Essen}, D.C.: \newblock Distributed hierarchical processing in the primate cerebral cortex. \newblock Cereb. Cortex \textbf{1} (1991)  1--47
\bibitem{Potjans14_785} Potjans, T.C., Diesmann, M.: \newblock The cell-type specific cortical microcircuit: Relating structure and activity in a full-scale spiking network model. \newblock Cereb. Cortex \textbf{24}(3) (2014)  785--806
\bibitem{Albada15} van Albada, S.J., Helias, M., Diesmann, M.: \newblock Scalability of asynchronous networks is limited by one-to-one mapping   between effective connectivity and correlations. \newblock PLOS Comput. Biol. \textbf{11}(9) (2015)  e1004490
\bibitem{Rosvall09} Rosvall, M., Axelsson, D., Bergstrom, C.T.: \newblock The map equation. \newblock Eur. Phys. J. Spec. Top. \textbf{178}(1) (2009)  13--23
\bibitem{Markov14} Markov, N.T., Vezoli, J., Chameau, P., Falchier, A., Quilodran, R., Huissoud,   C., Lamy, C., Misery, P., Giroud, P., Ullman, S., Barone, P., Dehay, C.,   Knoblauch, K., Kennedy, H.: \newblock Anatomy of hierarchy: Feedforward and feedback pathways in macaque visual cortex. \newblock J. Compar. Neurol. \textbf{522}(1) (2014)  225--259
\bibitem{Markov11_1254} Markov, N.T., Misery, P., Falchier, A., Lamy, C., Vezoli, J., Quilodran, R.,   Gariel, M.A., Giroud, P., Ercsey-Ravasz, M., Pilaz, L.J., Huissoud, C.,   Barone, P., Dehay, C., Toroczkai, Z., {Van Essen}, D.C., Kennedy, H.,   Knoblauch, K.: \newblock Weight consistency specifies regularities of macaque cortical networks. \newblock Cereb. Cortex \textbf{21}(6) (2011)  1254--1272
\bibitem{Markov2014_17} Markov, N.T., Ercsey-Ravasz, M.M., Ribeiro~Gomes, A.R., Lamy, C., Magrou, L.,   Vezoli, J., Misery, P., Falchier, A., Quilodran, R., Gariel, M.A., Sallet,   J., Gamanut, R., Huissoud, C., Clavagnier, S., Giroud, P., Sappey-Marinier,   D., Barone, P., Dehay, C., Toroczkai, Z., Knoblauch, K., Van~Essen, D.C.,   Kennedy, H.: \newblock A weighted and directed interareal connectivity matrix for macaque   cerebral cortex. \newblock Cereb. Cortex \textbf{24}(1) (2014)  17--36
\bibitem{Hilgetag19_905} Hilgetag, C.C., Beul, S.F., van Albada, S.J., Goulas, A.: \newblock An architectonic type principle integrates macroscopic   cortico-cortical connections with intrinsic cortical circuits of the primate brain. \newblock Netw. Neurosci. \textbf{3}(4) (2019)  905--923
\bibitem{Schuecker17} Schuecker, J., Schmidt, M., van Albada, S.J., Diesmann, M., Helias, M.: \newblock Fundamental activity constraints lead to specific interpretations of   the connectome. \newblock PLOS Comput. Biol. \textbf{13}(2) (February 2017)  e1005179
\bibitem{Muller15_11} Muller, E., Bednar, J.A., Diesmann, M., Gewaltig, M.O., Hines, M., Davison,   A.P.: \newblock Python in neuroscience. \newblock Front. Neuroinform. \textbf{9} (April 2015) ~11
\bibitem{Koester12_2520} K{\"o}ster, J., Rahmann, S.: \newblock Snakemake---a scalable bioinformatics workflow engine. \newblock Bioinformatics \textbf{28}(19) (2012)  2520--2522
\bibitem{Oliphant07} Oliphant, T.E.: \newblock Guide to NumPy. \newblock Trelgol Publishing, USA (2006) http://numpy.scipy.org.
\bibitem{Nest2140} Peyser, A., Sinha, A., Vennemo, S.B., Ippen, T., Jordan, J., Graber, S.,   Morrison, A., Trensch, G., Fardet, T., M{\o}rk, H., Hahne, J., Schuecker, J.,   Schmidt, M., Kunkel, S., Dahmen, D., Eppler, J.M., Diaz, S., Terhorst, D.,   Deepu, R., Weidel, P., Kitayama, I., Mahmoudian, S., Kappel, D., Schulze, M.,   Appukuttan, S., Schumann, T., Tun\c{c}, H.C., Mitchell, J., Hoff, M.,   M\"{u}ller, E., Carvalho, M.M., Zajzon, B., Plesser, H.E.: \newblock NEST 2.14.0 (2017)
\bibitem{Evans2011_fb} Evans, J.: \newblock Scalable memory allocation using jemalloc. \newblock   \url{https://www.facebook.com/notes/facebook-engineering/scalable-memory-allocation-using-jemalloc/480222803919}   (2011)
\bibitem{Ippen2017_30} Ippen, T., Eppler, J.M., Plesser, H.E., Diesmann, M.: \newblock Constructing neuronal network models in massively parallel   environments. \newblock Front. Neuroinform. \textbf{11} (2017) ~30
\bibitem{Brunel00} Brunel, N.: \newblock Dynamics of sparsely connected networks of excitatory and inhibitory spiking neurons. \newblock J. Comput. Neurosci. \textbf{8}(3) (2000)  183--208
\bibitem{Helias12_26} Helias, M., Kunkel, S., Masumoto, G., Igarashi, J., Eppler, J.M., Ishii, S.,   Fukai, T., Morrison, A., Diesmann, M.: \newblock Supercomputers ready for use as discovery machines for neuroscience. \newblock Front. Neuroinform. \textbf{6} (2012) ~26
\bibitem{Kunkel14_78} Kunkel, S., Schmidt, M., Eppler, J.M., Masumoto, G., Igarashi, J., Ishii, S.,   Fukai, T., Morrison, A., Diesmann, M., Helias, M.: \newblock Spiking network simulation code for petascale computers. \newblock Front. Neuroinform. \textbf{8} (2014) ~78
\bibitem{Kunkel2017_11} Kunkel, S., Schenck, W.: \newblock The nest dry-run mode: Efficient dynamic analysis of neuronal network simulation code. \newblock Front. Neuroinform. \textbf{11} (2017) ~40
\bibitem{Kunkel2012_5_35} Kunkel, S., Potjans, T.C., Eppler, J.M., Plesser, H.E., Morrison, A., Diesmann,   M.: \newblock Meeting the memory challenges of brain-scale simulation. \newblock Front. Neuroinform. \textbf{5} (2012) ~35
\bibitem{Morrison05a} Morrison, A., Mehring, C., Geisel, T., Aertsen, A., Diesmann, M.: \newblock Advancing the boundaries of high connectivity network simulation with   distributed computing. \newblock Neural Comput. \textbf{17}(8) (2005)  1776--1801
\bibitem{Eliasmith14_1} Eliasmith, C., Trujillo, O.: \newblock The use and abuse of large-scale brain models. \newblock Curr. Opin. Neurobiol. \textbf{25} (2014)  1--6
\bibitem{Fregnac17_6362} Fr{\'e}gnac, Y.: \newblock Big data and the industrialization of neuroscience: A safe roadmap for understanding the brain? \newblock Science \textbf{358}(6362) (2017)  470--477
\bibitem{Bassett18_566} Bassett, D.S., Zurn, P., Gold, J.I.: \newblock On the nature and use of models in network neuroscience. \newblock Nat. Rev. Neurosci. \textbf{19}(9) (2018)  566--578
\bibitem{Einevoll19_735} Einevoll, G.T., Destexhe, A., Diesmann, M., Gr{\"u}n, S., Jirsa, V., de~Kamps,   M., Migliore, M., Ness, T.V., Plesser, H.E., Sch{\"u}rmann, F.: \newblock The scientific case for brain simulations. \newblock Neuron \textbf{102}(4) (2019)  735--744
\bibitem{Proix17_641} Proix, T., Bartolomei, F., Guye, M., Jirsa, V.K.: \newblock Individual brain structure and modelling predict seizure propagation. \newblock Brain \textbf{140}(3) (2017)  641--654
\bibitem{Senk17_243} Senk, J., Yegenoglu, A., Amblet, O., Brukau, Y., Davison, A., Lester, D.R.,   L\"{u}hrs, A., Quaglio, P., Rostami, V., Rowley, A., Schuller, B., Stokes,   A.B., van Albada, S.J., Zielasko, D., Diesmann, M., Weyers, B., Denker, M.,   Gr\"{u}n, S.: \newblock A collaborative simulation-analysis workflow for computational   neuroscience using {HPC}. \newblock In {Di Napoli}, E., Hermanns, M.A., Iliev, H., Lintermann, A., Peyser, A., eds.: High-Performance Scientific Computing. {JHPCS} 2016. Volume 10164 of Lecture Notes in Computer Science., Springer, Cham (2017)  243--256
\bibitem{Linden11_859} Lind\'{e}n, H., Tetzlaff, T., Potjans, T.C., Pettersen, K.H., Gr\"{u}n, S.,   Diesmann, M., Einevoll, G.T.: \newblock Modeling the spatial reach of the {LFP}. \newblock Neuron \textbf{72}(5) (2011)  859--872
\bibitem{Cain16_e1005045} Cain, N., Iyer, R., Koch, C., Mihalas, S.: \newblock The computational properties of a simplified cortical column model. \newblock PLOS Comput. Biol. \textbf{12}(9) (2016)  e1005045
\bibitem{Hagen16} Hagen, E., Dahmen, D., Stavrinou, M.L., Lind\'{e}n, H., Tetzlaff, T., van Albada, S.J., Gr\"{u}n, S., Diesmann, M., Einevoll, G.T.: \newblock Hybrid scheme for modeling local field potentials from point-neuron networks. \newblock Cereb. Cortex \textbf{26}(12) (2016)  4461--4496
\bibitem{Schwalger17_e1005507} Schwalger, T., Deger, M., Gerstner, W.: \newblock Towards a theory of cortical columns: From spiking neurons to   interacting neural populations of finite size. \newblock PLOS Comput. Biol. \textbf{13}(4) (2017)  e1005507
\bibitem{Shimoura18_ReScience} Shimoura, R.O., Kamiji, N.L., de~Oliveira~Pena, R.F., Cordeiro, V.L., Ceballos,   C.C., Romaro, C., Roque, A.C.: \newblock {[Re]} the cell-type specific cortical microcircuit: relating structure and activity in a full-scale spiking network model. \newblock ReScience \textbf{4}(1) (2018)
\bibitem{VanAlbada18_291} van Albada, S.J., Rowley, A.G., Senk, J., Hopkins, M., Schmidt, M., Stokes,   A.B., Lester, D.R., Diesmann, M., Furber, S.B.: \newblock Performance comparison of the digital neuromorphic hardware {SpiNNaker} and the neural network simulation software {NEST} for a   full-scale cortical microcircuit model. \newblock Front. Neurosci. \textbf{12} (2018)  291
\bibitem{Knight18_941} Knight, J.C., Nowotny, T.: \newblock {GPUs} outperform current {HPC} and neuromorphic solutions in terms of speed and energy when simulating a highly-connected cortical model. \newblock Front. Neurosci. \textbf{12} (2018)  941
\bibitem{Rhodes19_20190160} Rhodes, O., Peres, L., Rowley, A.G.D., Gait, A., Plana, L.A., Brenninkmeijer, C.,   Furber, S.B.: \newblock Real-time cortical simulation on neuromorphic hardware. \newblock Phil. Trans. R. Soc. A \textbf{378} (2019)  20190160
\bibitem{Knight2020_bioRxiv} Knight, J.C., Nowotny, T.: \newblock Larger GPU-accelerated brain simulations with procedural connectivity. \newblock BioRxiv (2020)
\end{thebibliography}
\end{document}